\documentstyle[12pt,epsfig]{article}
\newlength{\dinwidth}
\newlength{\dinmargin}
\def\bq{\begin{equation}}
\def\eq{\end{equation}}
\def\bqa{\begin{eqnarray}}
\def\eqa{\end{eqnarray}}
\def\bqb{\begin{eqnarray*}}
\def\eqb{\end{eqnarray*}}
\def\to{\rightarrow}

\def\sc{\widetilde{c}}

\def\st{\widetilde{t}}
\def\sb{\widetilde{b}}

\def\sz{{\widetilde{\chi}}^0}
\def\sz1{{\widetilde{\chi}}^0_{1}}

\def\msz1{m_{\sz1}}

\def\tanbe{\tan\beta}
\def\nle{{\stackrel{<}{\sim}}}
\def\nge{{\stackrel{>}{\sim}}}

\def\stl{\st_{1}}

\def\sth{\st_{2}}
\def\sbl{\sb_{1}}

\def\sbh{\sb_{2}}

\def\mstl{m_{\stl}}
\def\msth{m_{\sth}}

\def\msbl{m_{\sbl}}

\def\tht{\theta_{t}}
\def\thb{\theta_{b}}

\def\mtil{\widetilde{m}}

\def\rb{\slash\hspace{-7pt}R}

\begin{document}
~~~\\
\vspace{10mm}
\begin{flushright}
AP-SU-00/01  \\
hep-ph/0007200 \\
July, 2000
\end{flushright}
\begin{center}
  \begin{Large}
   \begin{bf}
Single Sbottom/Scharm Production at HERA in an $R$-Parity Breaking Supersymmetric 
Model\\
   \end{bf}
  \end{Large}
  \vspace{5mm}
  \begin{large}
Tadashi Kon \\
  \end{large}
Faculty of Engineering, Seikei University, Tokyo 180-8633, Japan \\
kon@apm.seikei.ac.jp\\
 \vspace{3mm}
 \begin{large}
    Tetsuro Kobayashi\\
  \end{large}
Faculty of Engineering, 
Fukui Institute of Technology, Fukui 910-8505, Japan \\
koba@ge.seikei.ac.jp\\
 \vspace{3mm}
 \begin{large}
    Shoichi Kitamura\\
  \end{large}
Tokyo Metropolitan University of Health Sciences, Tokyo 116-8551, Japan\\
    kitamura@post.metro-hs.ac.jp\\
 \vspace{3mm}
 \begin{large}
    Takashi Iimura\\
  \end{large}
Faculty of Engineering, Seikei University, Tokyo 180-8633, Japan \\
iimura@dirac.ap.seikei.ac.jp\\  
\vspace{5mm}
\end{center}
\vskip50pt
\begin{quotation}
\noindent
\begin{center}

{\bf Abstract}
\end{center}
We investigate the production process of the single scalar bottom quark 
(sbottom) and scalar charm quark (scharm) at HERA. 
The sbottom and scharm could be produced via an $R$-parity breaking 
interactions $\lambda'_{123} \neq 0$ 
in the Minimal Supersymmetric Standard Model (MSSM). 
These processes give a slight excess in the invariant mass 
$M_{eq}$ distribution of the high $Q^2$ deep inelastic scattering for 
sufficiently heavy gauginos and the scalar top (stop).  
For the light gauginos and stop, 
it is shown that the tagging of $b/c$-quarks and charged leptons 
with high transverse energies will be indispensable to search for 
the processes.
\end{quotation}
\vfill\eject
\section{\it Introduction}
\renewcommand{\thefootnote}{\fnsymbol{footnote}}
The HERA accelerator is the only existing electron 
\footnote{In this paper the term "electron" is used to describe generically 
electrons or positrons if not specified.}
-- proton collider with the center of mass energy of 300GeV 
(318GeV since 1998) produced by 27.5GeV electrons and 820GeV 
(920GeV since 1998) protons \cite{schedule}. 
The luminosity has steadily increased and the integrated luminosity of 
$e^+p$ collisions during 1994 -- 1997 has reached 36.5pb$^{-1}$ and 
47.7pb$^{-1}$ for H1 and ZEUS, respectively. 
One of the most attractive features of the HERA is to provide a unique 
possibility to search for new phenomena revealing the physics beyond the 
standard model (SM) through rare topologies in the final states characteristic 
to electron -- proton collisions. 

In this paper we investigate a single scalar bottom quark (sbottom) and a scalar 
charm quark (scharm) production which is feasible at HERA 
$ep \to \sb\,X (\sc\, X)$. 
Our theoretical framework is on the basis of the Minimal Supersymmetric (SUSY) 
Standard Model (MSSM) with $R$-parity breaking (RB) interactions \cite{rbphys}. 

  In 1997 both the H1 \cite{H1} and ZEUS \cite{ZEUS} collaborations 
  reported an event excess 
in comparison with the SM expectations in the deep inelastic 
scattering (DIS) $e^+p \to e^+ X$ at large $x$ and high $Q^2$. 
The news very much excited high energy physics comminity. 
Various possibilities to understand the anomaly 
have extensively been examined by theoreticians since then 
\cite{hewett,dreiner,rbphys}. 
We have also proposed an interpretation of the anomalous event by the scalar top quark 
(stop) production in the SUSY models with RB interactions \cite{stoprb,dbst}. 
Contrary to initial expectation, the novel features have gradually faded 
in the whole data sample with increasing experimental data. 
However, this fact shows that HERA could have the potentiality exploring 
physics beyond the SM. 

Here we are concerned with 
$ep \to \sb/\sc\,X$ in contrast with $ep \to \st\,X$ in a previous work. 
The experimental signature of the present process has a large variety 
corresponding to decay modes of the sbottom/scharm. 
When decay modes $\sb\to ec$ and $\sc\to eb$ are dominant, 
the slight excess will be understood by the present scenario. 
On the other hand, if the gaugino modes dominate over previous ones, 
high $P_T$ lepton(s) and $b/c$-quarks would be typical signatures 
of our processes. 
In this respect, we are interested in the high $P_T$ leptons observed 
by the H1 \cite{highptl}. 

Throughout the present work 
whole calculations of decay widths and cross sections 
have been  performed by using the {\tt GRACE-SUSY} system, 
an automatic computation 
program for SUSY processes \cite{sgrace}. 
While the {\tt GRACE-SUSY} system is originally designed to treat such elementary 
sub-processes as $e^+e^-$, $eq$ and $qq$ collisions, 
we have recently succeeded in extending to $ep$ and $pp$ collisions. 
We use the extended new versions which 
includes an interface to the PDFLIB too. 
For the parton distribution function we have used CTEQ4M \cite{cteq}. 

\section{\it Models and constraints}

We start from the interaction Lagrangian 
based on the MSSM with an RB interaction  
\begin{equation}
L=\lambda'_{1jk} ({\widetilde{u_{jL}}} {\overline{d_k}} P_L e 
- {\overline{\widetilde{d_{kR}}}} \overline{e^c} P_L u_j) + h.c,  
\label{sqRb}
\end{equation}
where $P_{L,R}$ read left and right  handed chiral projection operators, 
respectively.   
The interaction Lagrangian (\ref{sqRb}) has been derived from the general RB 
superpotential  \cite{Barger}; 
\begin{equation}
W_{\rb}=\lambda_{ijk}\hat{L}_i \hat{L}_j \hat{E^c}_k 
+ \lambda'_{ijk}\hat{L}_i \hat{Q}_j \hat{D^c}_k + 
\lambda''_{ijk}\hat{U^c}_i \hat{D^c}_j \hat{D^c}_k, 
\label{RBW}
\end{equation}
where $i$, $j$ and $k$ are generation indices. 
The first two terms violate the lepton number $L$ and the last term 
violates the baryon number $B$. 
Incorporating RB interactions into the MSSM 
we have a possibility to unveil yet unresolved problems as 
({\romannumeral 1}) the cosmic baryon number violation, 
({\romannumeral 2}) the origin of the masses and the 
magnetic moments of neutrinos and 
({\romannumeral 3}) some interesting rare processes 
induced by the $L$ and/or $B$ violation. 
The realization of the coupling among participating particles 
in Eq.~(\ref{sqRb}) 
will be most suitable for the $ep$ collider HERA 
because the squark ${\widetilde{u_{jL}}}$ or ${\overline{\widetilde{d_{kR}}}}$ 
will be produced in the $s$-channel 
in $e^+$-$q$ sub-processes. 
\begin{eqnarray}
&& e^+ + d_k \to {\widetilde{u_{jL}}}\\
&& e^+ + {\overline{u_j}} \to {\overline{\widetilde{d_{kR}}}}.
\end{eqnarray}
The upper bounds on the coupling constants $\lambda'_{1jk}$ have already been 
settled by, for instance, neutrinoless double beta decay \cite{beta,mohap}, 
charged current universality \cite{Barger,ccu}, atomic parity violation (APV) 
\cite{Barger,ccu,apv} and $\nu_e$ mass \cite{nemass}. 
Some of possible nine coupling constants $\lambda'_{1jk}$ are severely 
constrained by experiments mentioned above. 

Here we pay attention to the single sbottom and scharm production, 
\begin{eqnarray}
&& e^+ + {\overline{c}} \to {\overline{\widetilde{b_{R}}}} \\
&& e^+ + b \to {\widetilde{c_{L}}},
\end{eqnarray}
which are realized via a non-zero $R$-parity breaking coupling 
$\lambda'_{123}$. 
The most stringent upper bound on $\lambda'_{123}$ comes  
from the charged current universality experiments \cite{Barger}, 
\begin{equation}
\lambda'_{123} \nle 0.1
\end{equation}
for $m_{\sb_{R}} \simeq 200$GeV.

Sbottoms (${\widetilde{b_L}}$, ${\widetilde{b_R}}$)
as well as stops are naturally mixed each 
other due to a large Yukawa coupling to their partner quark. 
The left- and right-handed sbottom 
(${\widetilde{b_L}}$, ${\widetilde{b_R}}$) 
are expressed in terms of 
the mass eigenstates ($\sbl$, $\sbh$) through a mixing angle $\thb$. 
\begin{eqnarray}
{\widetilde{b_L}} &=& \sbl\cos\thb - \sbh\sin\thb \\
{\widetilde{b_R}} &=& \sbl\sin\thb + \sbh\cos\thb.
\end{eqnarray}
We should note that the sbottom mixing is considered to be natural in the 
case of large $\tan\beta$, because the off-diagonal terms of the sbottom 
mass matrix are proportional to $m_b (A_b+\mu \tan\beta)$, where 
$A_b$ and $\mu$, respectively, denote the trilinear coupling and 
the SUSY Higgs mass. 
Then the interaction Lagrangian (\ref{sqRb}) can easily be rewritten 
in terms of the mass eigenstates (${\widetilde{b_1}}$, ${\widetilde{b_2}}$). 
In particular, it is worthy to note that both sbottoms 
$\sbl$ and $\sbh$ could be 
produced in the $eq$ collisions through 
\begin{equation}
L_{\sb e c}=-\lambda'_{123} (\sin\thb {\overline{\widetilde{b_{1}}}} 
\overline{e^c} P_L c
                           + \cos\thb {\overline{\widetilde{b_{2}}}} 
\overline{e^c} P_L c) + h.c. . 
\label{stRb}
\end{equation}
We would emphasize that only one species of scharm ${\widetilde{c_L}}$ can  
couple to $e^+ b$ in our scenario with $\lambda'_{123}$ $\neq$ 0. 

Interestingly, there exists a theoretical upper bound on 
the mass of the lighter sbottom $\sbl$. 
In the MSSM, 
since the left handed stop $\st_L$ and sbottom $\sb_L$ form an $SU(2)$ 
doublet, their masses include the same contribution from a soft scalar breaking mass 
${\mtil_{Q_3}}$.  
We obtain an upper bound on the sbottom mass 
$m^2_{\sbl}$ as 
\begin{equation}
m^2_{\sbl}  \le m^2_{\sth} - m^2_t + m^2_b - m^2_W \cos{2\beta}.  
\label{upper}
\end{equation}
The lighter sbottom $\sbl$ cannot be heavy for relatively light $\sth$. 
As we consider the single sbottom and scharm production, we should take 
into account 
consistent model parameter sets allowing $m_{\sb}, m_{\sc}$ $\sim$ 200GeV. 
For example, in order to obtain $m_{\sbl}$ $\nge$ 200GeV, the heavier stop 
must be rather heavy $m_{\sth}$ $\nge$ 250GeV. 
This means that the lighter stop must be lighter than about 250GeV. 

Next we examine the decay modes of the sbottom and scharm. 
We assume that the sbottom can decay through 
$\sb_{1,2}$ $\to$ $e\,c$, $b\,{{\widetilde{\chi}}}^0_i$ and $W\,\stl$, 
where ${{\widetilde{\chi}}}^0_i$ denotes neutralinos ($i = 1 \sim 4$). 
On the other hand, the scharm can decay into 
$\sc_L$ $\to$ $e\,b$, $c\,{{\widetilde{\chi}}}^0_i$ and 
$s\,{{\widetilde{\chi}}}^+_{1,2}$. 
The first decay mode for both sbottom and scharm occurs via 
the $R$-parity breaking mode and the others proceed through 
$R$-parity conserving interactions.

At present, we know that 
the most stringent mass bound on the sbottom with the dominant 
decay mode $\sbl \to ec$ is 
$\msbl > 242 (204)$GeV for Br($\sbl \to ec$)=1 (0.5).
This has been obtained by the combined experimental results of 
the leptoquark searches by the CDF and D0 experiments at Tevatron \cite{lqlim}. 
It should be noted that the mass bound could become lower when 
the branching ratio of the $R$-parity breaking mode $\sb_1$ $\to$ $e\,c$ 
is smaller.

We are aware of constraints from the precision measurements at LEP1. 
Potentially, contributions to the $\Delta\rho$ from the $\st_L$ and $\sb_L$ 
becomes 
large if their masses are not so different from the weak mass scale of $m_Z$. 
We have checked contributions 
to the $\Delta\rho$ from the $\st_i$ and $\sb_j$ \cite{deltarho} 
could become small to such extent as $7\times 10^{-4}$ for 
$\msbl =200$GeV, $\mstl = 100$GeV and $\msth = 250$GeV.

\section{Numrical results}

In Table {\uppercase\expandafter{\romannumeral 1}} we show two typical 
input parameter sets adopted throughout our 
numerical calculations. 
Output mass paramters of the neutralinos and the lighter 
chargino are also presented for reference. 

\begin{center}
{{\bf Table {\uppercase\expandafter{\romannumeral 1}}}} \quad 
{Typical input parameter sets}\\
\end{center}
\begin{center}
\begin{tabular}{|c|rr|}
\hline
masses in GeV  &     A    &     B \\
\hline
$M_2$   & $300$     & $150$     \\
$\tanbe$& $12$   & $2$   \\
$\mu$   & $-300$   & $-300$ \\
$m_{\widetilde{c_{L}}}$  & $235$   & $230$  \\
$m_{\widetilde{t_{1}}}$  & $250$   & $ 90$  \\
$m_{\widetilde{t_{2}}}$  & $400$   & $300$  \\
$\tht$  & $0.42$  & $1.0$  \\
$\thb$  & $1.2$  & $1.0$  \\
\hline
$m_{\widetilde{\chi_{1}}^0}$  & $143.4$   & $76.2$  \\
$m_{\widetilde{\chi_{2}}^0}$  & $254.4$   & $158.3$  \\
$m_{\widetilde{\chi_{1}}^+}$  & $254.4$   & $158.4$  \\
\hline
\end{tabular}
\label{table1}
\end{center}

For these parameter sets, we first calculate the decay branchig ratios 
of sbottoms and the left-handed scharm 
(see Table {\uppercase\expandafter{\romannumeral 2}}), 
where we take $\lambda'_{123}=0.2 [0.1]$ and 
$m_{\widetilde{b_{1}}}=225 [200]$GeV for 
the set(A) [(B)]. 

\begin{center}
{{\bf Table {\uppercase\expandafter{\romannumeral 2}}}} \quad 
{Branching ratios}\\
\end{center}
\begin{center}
\begin{tabular}{|c|rr|}
\hline
      &     A    &     B \\
\hline
$\sb_1 \to e\,c$   & $0.688$     & $0.084$     \\
$\sb_1 \to b\,{{\widetilde{\chi}}^0}_1$   & $0.312$     & $0.334$     \\
$\sb_1 \to b\,{{\widetilde{\chi}}^0}_2$   & $0.0$     & $0.181$     \\
$\sb_1 \to W\,{{\widetilde{t}}}_1$   & $0.0$     & $0.401$     \\
$\sb_1$ total width & 0.090GeV  &  0.164GeV    \\
\hline
$\sb_2 \to e\,c$   & $0.693$     & $0.075$     \\
$\sb_2 \to b\,{{\widetilde{\chi}}}^0_1$   & $0.307$     & $0.095$     \\
$\sb_2 \to b\,{{\widetilde{\chi}}}^0_2$   & $0.0$     & $0.278$     \\
$\sb_2 \to W\,{{\widetilde{t}}}_1$   & $0.0$     & $0.551$     \\
$\sb_2$ total width & 0.179GeV  &  0.352GeV    \\
\hline
$\sc_L \to e\,b$   & $0.959$     &   $0.006$   \\
$\sc_L \to c\,{{\widetilde{\chi}}}^0_1$   & $0.041$     &  $0.069$   \\
$\sc_L \to c\,{{\widetilde{\chi}}}^0_2$   & $0.0$       &  $0.291$   \\
$\sc_L \to s\,{{\widetilde{\chi}}}^+_1$   & $0.0$       &  $0.634$   \\
$\sc_L$ total width & 0.191GeV  &   $0.697$  \\
\hline
\end{tabular}
\label{table2}
\end{center}

From Tables {\uppercase\expandafter{\romannumeral 1}} and 
{\uppercase\expandafter{\romannumeral 2}}, 
we find that the set(A) is characterized by the dominant decay modes  
$\sb_{1,2} \to e\,c$ and $\sc_L \to e\,b$. 
On the other hand, the SUSY decay modes 
$\sb_{1,2} \to b\,{{\widetilde{\chi}}^0}_i$ and 
$\sb_1 \to W\,{{\widetilde{t}}}_1$ are dominated in the set(B). 
These properties are originated from the different values of 
the SU(2) gaugino SUSY breaking mass $M_2$ and 
the lighter stop mass $m_{\widetilde{t_{1}}}$. 
Large $M_2$ in the set(A) corresponds to heavy neutralinos 
${\widetilde{\chi_{i}}^0}$. 
The $R$-pariry conserving decay modes of the sbottoms/scharm are 
kinematically forbidden in this case.

\begin{figure}[hbtp]
\centerline{\epsfig{figure=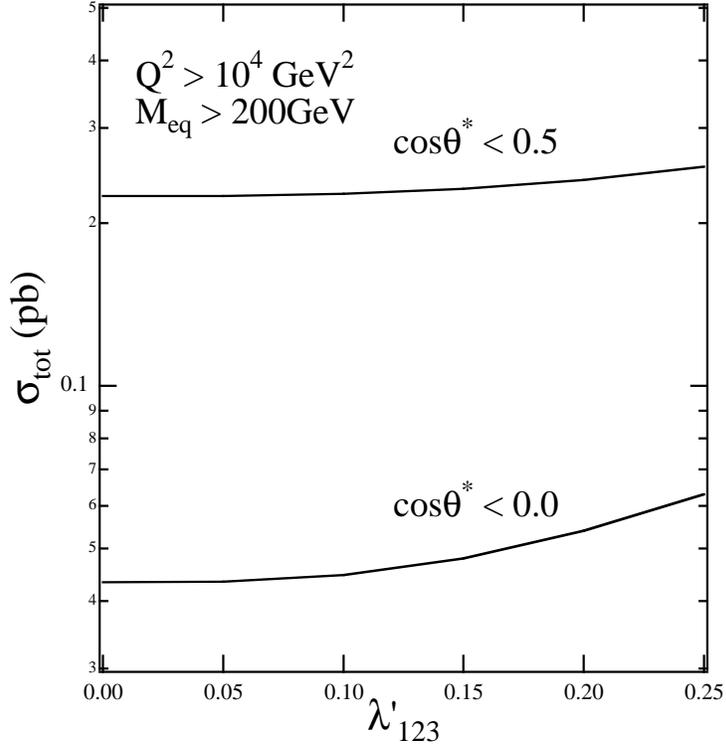,height=10cm,angle=0}}
 \caption{ $\lambda'_{123}$ dependence of the total cross section 
 for $e^+ p \to e^+ q X$. We adopt parameter set(A) and 
 $m_{\widetilde{b_{1}}}=225$GeV.
 \label{fig:tot}}
\end{figure}
In Fig.1 we show the $\lambda'_{123}$ dependence of the total cross section 
for the process, 
\begin{equation}
e^+ p \to e^+ q X, 
\end{equation}
where we take the set(A) and $m_{\widetilde{b_{1}}}=225$GeV is assumed. 
To extract the relevant signal we adopt kinematical cuts 
$Q^2 > 10^4$GeV$^2$ and $M_{eq} > 200$GeV. 
In our calculation two kinds of cuts on $\theta^*$, the angle between 
the outgoing and incoming positron in the $eq$ rest frame, are 
introduced. 
As clearly seen from Fig.1, the cross section is more sensitive to the 
$\lambda'_{123}$ for the more restrictive cut $\cos\theta^* <0$. 
In another words, more restrictive cut on $\cos\theta^*$ would be efficient to 
extract the signal from background. 

\begin{figure}[hbtp]
\centerline{\epsfig{figure=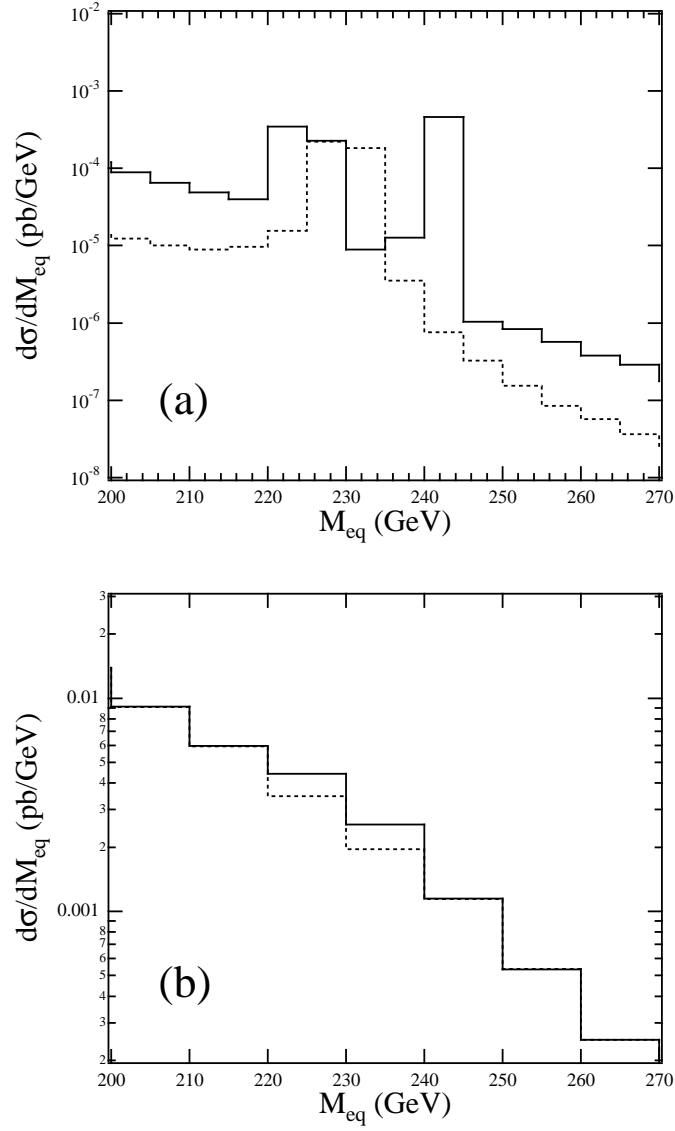,height=15cm,angle=0}}
 \caption{ $M_{eq}$ distribution for set(A), 
 $\lambda'_{123}=0.2$, $m_{\widetilde{b_{1}}}=225$GeV 
 $Q^2 > 10^4$GeV$^2$ and $\cos\theta^* < 0.5$. 
 (a) $e^+ p \to e^+ \overline{c} X$ (solid line) and 
 $e^+ p \to e^+ b X$ (dotted line) including contributions from 
 $\widetilde{b_{1,2}}$ and $\widetilde{c_{L}}$, respectively. 
 (b) the signal (solid line) and the SM background (dotted line) 
 for $e^+ p \to e^+ q X$. 
 \label{fig:mqdist}}
\end{figure}
In Fig.2 we show the differential cross section against 
$M_{eq}$ for the set(A),  
$\lambda'_{123}=0.2$, $m_{\widetilde{b_{1}}}=225$GeV, 
$Q^2 > 10^4$GeV$^2$ and $\cos\theta^* < 0.5$. 
It will be seen from 
Fig.2(a) that only the sbottom/scharm production contributes to 
the relevant signal. 
As the masses of two sbottoms and left-handed scharm are in the set(A) 
almost degenerate, three peaks corresponding to their masses 
appear in the range of $M_{eq} =$ 220 $\sim$ 240GeV. 
Summing up all sub-process contributions we obtain the 
observable differential cross section represented by 
Fig.2(b). 
We can clearly see a slight excess of the cross section 
in the range of $M_{eq}=$ 220 $\sim$ 240GeV. 
Apparent peak structures, however, cannot be identified. 

\begin{figure}[hbtp]
\centerline{\epsfig{figure=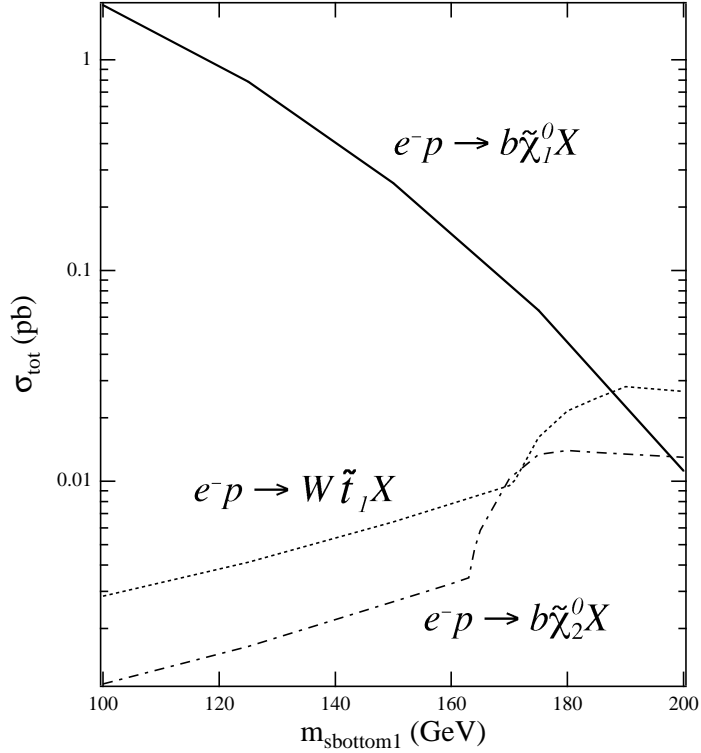,height=10cm,angle=0}}
 \caption{ The sbottom mass $m_{\widetilde{b_1}}$ dependence of the 
 sbottom total production cross 
sections with different final states. 
The set(B) and $\lambda'_{123}=0.1$ are assumed. 
 \label{fig:msbdep}}
\end{figure}
The mass $m_{\widetilde{b_1}}$ dependence of the total production 
cross sections of the sbottom with different final states is shown 
in Fig.3, where the set(B) and $\lambda'_{123}=0.1$ have been adopted. 
We should note that Br($\sbl \to ec$) $\nle$ 0.1 and even the light 
sbottom $m_{\widetilde{b_1}} \nge 100$GeV is not excluded from the leptoquark 
searches at Tevatron. 
Then 
signatures relevant to the sbottom production have a large variety of 
characteristic features. 
Relatively light sbottom $m_{\widetilde{b_1}} \nge 150$GeV dominantly decays into 
$b\,{{\widetilde{\chi}}}^0_1$ and the heavier sbottom also decays into 
$W\,{{\widetilde{t}}}_1$ and $b\,{{\widetilde{\chi}}}^0_2$. 
As the lightest neutralino ${{\widetilde{\chi}}}^0_1$ can decay into 
the ordinary particles $e^\pm c b$ or $b s \nu_e$, the decay chains leading to 
relevant signatures are written as follows,
\begin{eqnarray}
{\widetilde{b_1}} X &\to& b {{\widetilde{\chi}}}^0_1 X \to b (e^\pm c b) X,  
b (b s \nu_e) X \\
&\to& b {{\widetilde{\chi}}}^0_2 X \to b (\ell^+ \ell^- {{\widetilde{\chi}}}^0_1) X 
\to b (\ell^+ \ell^- (e^\pm c b)) X, b (\ell^+ \ell^- (b s \nu_e)) X \\
&\to& b {{\widetilde{\chi}}}^0_2 X \to b (b {\bar{b}} {{\widetilde{\chi}}}^0_1) X 
\to b (b {\bar{b}} (e^\pm c b)) X, b (b {\bar{b}} (b s \nu_e)) X \\
&\to& W {{\widetilde{t}}}_1 X \to (\ell \nu_\ell) (c {{\widetilde{\chi}}}^0_1) X 
\to (\ell \nu_\ell) (c (e^\pm c b)) X, (\ell \nu_\ell) (c (b s \nu_e)) X. 
\end{eqnarray}
High $P_T$ charged lepton(s) and/or $b/c$-jet(s) are typical signatures 
of these processes. 
Especially, multi-charged leptons and $b/c$-jets are almost the SM background 
free signals and they could effectively serve us in the sbottom search. 

\begin{figure}[hbtp]
\centerline{\epsfig{figure=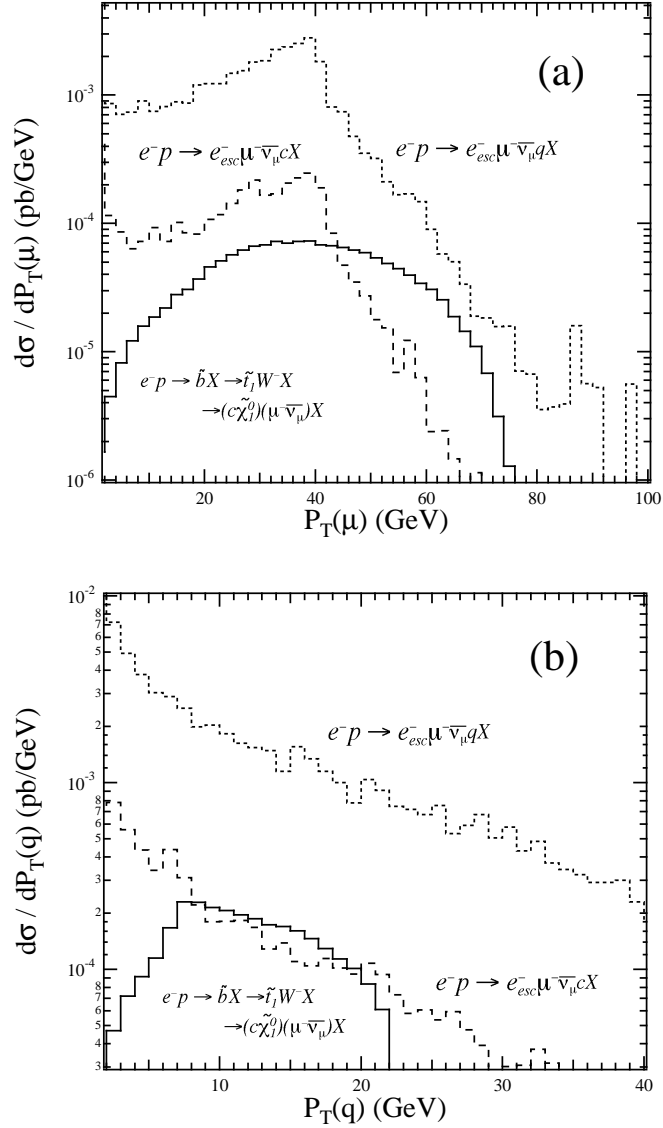,height=15cm,angle=0}}
 \caption{ Transverse momentum distributions of the final muon $P_T(\mu)$ (a) 
and the $c$-quark $P_T(q)$ (b) for the process 
$e^-p \to {\widetilde{b}} X \to W^- {{\widetilde{t}}}_1 X 
\to (\mu^- \nu_\mu) (c {{\widetilde{\chi}}}^0_1) X$ (solid line). 
The set(B), $m_{\widetilde{b_1}}=200$GeV and $\lambda'_{123}=0.1$ are assumed. 
Dotted and dashed lines, respectively, correspond to the backgrounds 
$e^-p \to e_{esc}W^-qX \to e_{esc}(\mu^-\nu_\mu)qX$ and 
$e^-p \to e_{esc}W^-cX \to e_{esc}(\mu^-\nu_\mu)cX$. 
 \label{fig:muqdist}}
\end{figure}
In Fig.4 
we show the differential cross sections for 
transverse momentum of the final muon $P_T(\mu)$ 
and the $c$-quark $P_T(q)$ for the process (16) as an example. 
Here we take the set(B), $m_{\widetilde{b_1}}=200$GeV and $\lambda'_{123}=0.1$. 
When we use the inclusive cross section of the muon to extract the signal, 
the single $W$ production, $ep \to e_{esc}WqX \to e_{esc}(\mu\nu_\mu)qX$, is  
severe background, where $e_{esc}$ means the final electron escaped into the beam pipe. 
As seen from Fig.4, the background cross section is much larger than the signal. 
However, if we can tag the $c$-quark, the background should be restricted to 
$ep \to e_{esc}WcX \to e_{esc}(\mu\nu_\mu)cX$ and its cross section is 
comparable to the signal. 
In fact we can expect the event excess in both $P_T(\mu)$ and $P_T(c)$ distributions. 
In addition to the muon and the $c$-quark, the information 
of the $b$-quark and/or the high $P_T$ electrons (positrons) 
coming from the neutralino ${{\widetilde{\chi}}}^0_1$ decays 
is available for us. 
All these informations make easier to extract the signals which 
we are searching for. 

\section{\it Concluding remarks}

We have investigated a possible scenario to understand  
an excess of large $x$ and high $Q^2$ events in the $e^+p \to e^+ X$ observed 
by the H1 and ZEUS experiments at HERA. 
Our reasoning is based upon the resonance production of 
the sbottoms and the scharm with an 
$R$-parity breaking 
interaction in the framework of the MSSM. 
We have focussed our attention upon its broad mass $M_{eq}$ distribution 
characteristic of the HERA events in addition to large $x$ and high $Q^2$. 
Assuming almost degenerate three mass eigenstates 
$\widetilde{b_1}$, $\widetilde{b_2}$ and $\widetilde{c_L}$, 
we have simulated the mass distribution on the basis of our specific scenario. 
 In the case of large $M_2$ and $m_{\widetilde{t_{1}}}$ the present scenario has 
 its validity.
 If this is not the case, the sbottom decay dominantly into lighter sparticles. 
 Then signals are characterized by high $P_T$ charged lepton(s) and/or 
 $b/c$-jet(s). 
Now planning vertex detectors at the H1 and ZEUS provide us the efficient $b/c$ tagging. 
Together with a luminosity upgrade bringing integrated luminosities, 
for instance, to the level of 1fb$^{-1}$, it is expected that an event excess 
will be confirmed and our MSSM approach opens a new horizon. 

\begin{flushleft}
{\Large{\bf Acknowledgements}}
\end{flushleft}
This work was supported in part by 
the Grant-in-Aid for Scientific Research from the Ministry of Education, 
Science and Culture of Japan, No. 10440080.



\end{document}